# Tapered Photonic Switching

**AUTHORS:** Emanuele Galiffi[1], Shixiong Yin[1], Andrea Alù[1,2*]

**AFFILIATIONS:**

[1]Photonics Initiative, Advanced Science Research Center, City University of New York, New York, NY 10031, USA (aalu@gc.cuny.edu)

[2]Physics Program, Graduate Center, City University of New York, New York, NY 10016, USA

*Corresponding author: e-mail: aalu@gc.cuny.edu

**ABSTRACT**

The advent of novel nonlinear materials has stirred unprecedented interest in exploring the use of temporal inhomogeneities to achieve novel forms of wave control, amidst the greater vision of engineering metamaterials across both space and time. When the properties of an unbounded medium are abruptly switched in time, propagating waves are efficiently converted to different frequencies, and partially coupled to their back-propagating phase-conjugate partners, through a process called time-reversal. However, in realistic materials the switching time is necessarily finite, playing a central role in the resulting temporal scattering features. By identifying and leveraging the crucial role of electromagnetic momentum conservation in time-reversal processes, here we develop a very general analytical formalism to quantify time-reversal due to temporal inhomogeneities of arbitrary profile. Finally, we deploy our analytic theory to develop a formalism, analogous to spatial tapering theory, that enables the tailoring of a desired time-reversal spectral response, demonstrating its use for the realization of broadband frequency converters and filters.

# 1. INTRODUCTION

Scattering is a key feature of wave propagation, occurring ubiquitously at the spatial interface between two media with different properties. It consists in the generation of reflected waves in the originating medium, and transmitted waves into the second medium. The temporal analogue of a spatial interface arises when the properties of an unbounded medium are abruptly and uniformly switched in time, and the associated phenomena have been attracting significant recent interest across various wave platforms [1]-[2]. Such temporal interfaces exhibit remarkable differences compared to their spatial counterparts: in particular, frequency and energy are not conserved at these time boundaries, whereas momentum is. Coupling between positive and negative frequencies corresponds to time-reversal at these switching events, implying that a portion of the energy associated with waves initially propagating in one direction flips its propagation direction, travelling backwards while the wavevector is conserved [1]-[3]. Interference between transmitted and time-reversed waves at time-interfaces can enable highly exotic wave phenomena, at the basis of the field of time metamaterials [1],[4]-[8], including parametric amplification [9]-[10], temporal aiming [11], topological phenomena within photonic time-crystals [12] and in synthetic frequency dimensions [13], non-Hermitian effects such as nonreciprocal gain [14], spectral causality [15] and temporal parity-time symmetry [16], temporal Anderson localization [17]-[18], unitary energy transfer between resonators [19], as well as efficient [20] and broadband [21] absorbers, among several others [1]. Interesting opportunities for new physics can also be found at the interplay between temporal interfaces and material dispersion [22]-[23]. While time-reversal is at the core of several of these phenomena, temporally reflected waves may be undesirable in many applications, in particular in the context of efficient frequency conversion. Antireflection coatings based on temporal multilayers, mimicking their spatial counterparts, have indeed been recently introduced to minimize the energy trapped into time-reversed waves [24]-[26].

Most research work in this area has so far been assuming that time interfaces are abrupt, i.e., that the time required to switch the material properties is negligible compared to the wave dynamics. However, in any realistic scenario the material response cannot be considered instantaneous, and in several instances the finite width of a temporal interface may become comparable with the period of the propagating signals, especially as we operate at higher frequencies. In addition, more interesting phenomena are observed at time interfaces involving a large contrast of the material properties before and after the switching event, and a tradeoff between permittivity contrast and switching speed is naturally expected [27]. In this Letter, we analytically investigate temporal interfaces that follow a continuous evolution in time with arbitrary profile, and we deploy our formulation to unravel the unexplored opportunities arising when the material responses are not instantaneous. In particular, we demonstrate that the control of the temporal evolution may enable efficient frequency conversion in the temporal analogue of a Klopfenstein taper [28]. Whilst we restrict ourselves to Maxwell's equations, the principles we invoke here are general, and our results can be extended to other wave realms. Our findings illuminate the role of momentum conservation in temporal scattering, shedding new light on the duality between spatial reflection and time-reversal in realistic settings.

# 2. RESULTS

We are commonly used to writing Maxwell's equations in the frequency domain by assuming a harmonic time dependence, since in static linear media frequency is conserved across spatial interfaces. At time interfaces, on the contrary, spatial momentum is conserved and not frequency, therefore it is convenient to consider spatially harmonic $e^{ikz}$ fields in space, where

$k$ is the wavenumber and $z$ is the propagation coordinate. Under this assumption, the displacement field $D$ and magnetic induction $B$ in a generally time-varying homogeneous medium obey at any point in space the temporal analogue of the telegrapher's equations:

$$\frac{\partial B}{\partial t} = -Z(t)D$$
$$\frac{\partial D}{\partial t} = -Y(t)B \quad , \quad (1)$$

where $Z = ik/\varepsilon(t)$ and $Y = ik/\mu(t)$, and $\varepsilon$ and $\mu$ are the permittivity and permeability of the material. At a time interface, $D$ and $B$ are continuous [29], hence

$$D(t^+) = (T+R)D(t^-)$$
$$B(t^+) = Z_0(t)(T-R)B(t^-) \quad , \quad (2)$$

where $R$ and $T$ are the scattering coefficients respectively associated with the backward (time-reversed) and forward waves generated at the time interface, and $Z_0(t) = \sqrt{Z(t)/Y(t)}$ is the wave impedance. We can invert Eq. (2) to yield the ratio $R/T$ as a function of the local wave impedance $B/D$ at instant $t$:

$$\rho(t) = R(t)/T(t) = \frac{Z_0(t) - B(t)/D(t)}{Z_0(t) + B(t)/D(t)}. \quad (3)$$

Using this result, we can derive a general solution for $\rho(t)$ as a function of an arbitrary time-modulation profile of the material properties. Dividing the first of Eq. (1) by $B$ and the second by $D$ and taking their difference, we find

$$\frac{\partial}{\partial t}\left(\ln B/D\right) = -\frac{Z(t)}{(B/D)} + Y(t)(B/D). \quad (4)$$

Combining (3) and (4), after some algebra we obtain

$$\frac{\partial}{\partial t}\ln Z_0 - \frac{2}{1-\rho^2}\frac{\partial \rho}{\partial t} + \frac{4\gamma\rho}{1-\rho^2} = 0, \quad (5)$$

with $\gamma = \sqrt{ZY} = ik/\sqrt{\varepsilon\mu}$. Eq. (5) can be linearized assuming $\rho^2 \ll 1$, a safe assumption in realistic scenarios. This yields

$$\frac{\partial \rho}{\partial t} - 2\gamma(t)\rho = F(t), \quad (6)$$

with $F(t) = \frac{1}{2}\frac{\partial}{\partial t}\ln Z_0$. The general solution is

$$\rho(t) = \int_{-\infty}^{t} F(t')e^{i\Phi(t)}dt', \quad (7)$$

which defines the ratio $R/T$ of time-reversed over transmitted signals in time for arbitrary variations of the material properties through $Z_0(t)$, with $i\Phi(t) = 2\int_{-\infty}^{t} \gamma(t')dt' = 2ik\int_{-\infty}^{t}[\varepsilon(t')\mu(t')]^{-1/2} dt'$.

In contrast with the case of a spatial interface, here we are concerned with the ratio $\rho = R/T$ because at time-interfaces the energy is not conserved, so $|T|$ and $|R|$ can both become arbitrarily large [1]. Conservation of the total electromagnetic momentum **P** in the medium, which is ensured by translational invariance, allows us to derive the actual time-reversal and transmission magnitudes. Assuming, without loss of generality, that only forward propagating waves are initially present, the total momentum density before and after an arbitrary time variation going from $Z_1$ to $Z_2$ is $P_1 = Z_1|D_1|^2$ and $P_2 = Z_2(|T|^2 - |R|^2)|D_1|^2$, respectively. Conservation of momentum therefore requires

$$|T|^2 - |R|^2 = Z_1/Z_2. \qquad (8)$$

As a result, whilst $|T|$ and $|R|$ can both change arbitrarily, their difference must remain constant. Note that we made no assumption here on the temporal variation $Z_0(t)$, so this result holds for any form of temporal switching as long as it is carried uniformly across the spatial extent of the wave. Combining this result with Eq. (3) yields

$$|T|^2 = \frac{Z_1}{Z_2}\frac{1}{1-|\rho_2|^2}; |R|^2 = \frac{Z_1}{Z_2}\frac{|\rho_2|^2}{1-|\rho_2|^2}, \qquad (9)$$

where $\rho_2$ is the ratio $\rho(t)$ at the end of the switching process.

## 2.1 TEMPORAL SCATTERING FROM SIGMOIDAL STEP

In the case of a step-like switching profile of permittivity, modeled as a sigmoid function with rise time $\tau$ of the form $\varepsilon_s = (\varepsilon_1 + \frac{\delta\varepsilon}{2}) + \frac{\delta\varepsilon}{2}\tanh(t/\tau)$, and a static permeability $\mu = 1$, the phase in Eq. (7) can be explicitly written as

$$i\Phi_s(t) = \int^{t} \frac{2ik}{\sqrt{A+B\tanh(t'/\tau)}} dt'$$

$$= 2ik\tau \left[ \frac{\tanh^{-1}\frac{\sqrt{A+B\tanh t/\tau}}{\sqrt{A+B}}}{\sqrt{A+B}} - \frac{\tanh^{-1}\frac{\sqrt{A+B\tanh t/\tau}}{\sqrt{A-B}}}{\sqrt{A-B}} \right], \qquad (10)$$

where $A = \varepsilon_1 + \delta\varepsilon/2$, $B = \delta\varepsilon/2$.

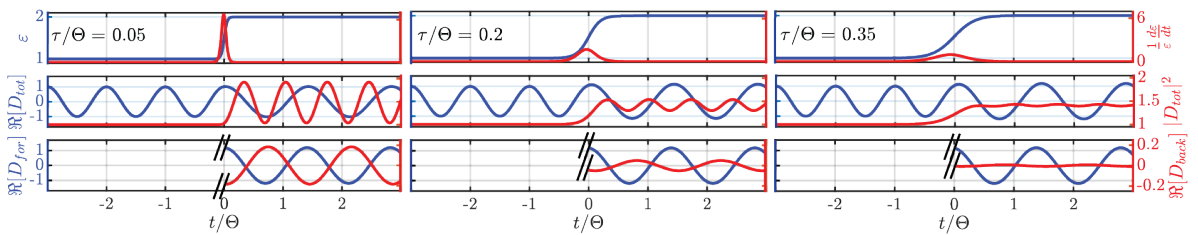

**Figure 1.** Top row: Permittivity profile $\varepsilon(t)$ and $\frac{1}{\varepsilon}\frac{d\varepsilon}{dt}$ for a sigmoidal step with rise time $\tau/\Theta = 0.05$ (left), $0.2$ (center) and $0.35$ (right), where $\Theta = 2\pi/\omega_1$ is the period of the incoming wave. The input frequency is $\omega_1 = 1$. Second row: total displacement field $\Re[D_{tot}]$ and squared modulus $|D|^2$ through the switching process. Third row: amplitude of the forward (blue, $\Re[Te^{(-i\omega(t \to \infty)t)}]$) and backward (red, $\Re[Re^{i\omega(t \to \infty)t}]$) wave at the end of the modulation process. In order to be able to demonstrate the sharp-step case with high accuracy, the results in this figure are computed numerically using the routine in [30]. The two methods are compared in Fig. 2.

Figure 1 shows (top) the permittivity profile $\varepsilon(t)$ (left axis, blue) and its normalized derivative $\frac{1}{\varepsilon}\frac{d\varepsilon}{dt}$, the time-evolution of the real part $\Re[D_{tot}]$ (second row, left axis) and squared modulus $|D|^2$ (second row, right axis) of the displacement field $D$ for three increasingly slower sigmoidal variations of permittivity $\varepsilon$ going from 1 to 2 with rise times $\tau = 0.05\Theta$, $0.2\Theta$ and $0.35\Theta$ (columns), where $\Theta = 2\pi/\omega_1$ is the period of the incoming wave. The bottom row shows the forward (blue) and backward (red) components of the displacement field $\Re[Te^{i\omega_2 t}]$ and $\Re[Re^{i\omega_2 t}]$ at the end of the switching process, where $\omega_2 = k/\sqrt{\varepsilon(t \to \infty)\mu_0}$ is the final angular frequency. The most evident effect of a slower transition is the amplitude decrease of the time-reversed wave, while the frequency-converted transmitted wave is preserved. In the sharp-step limit $\tau \to 0$, forward and backward waves have exactly opposite phases (assuming $\varepsilon_2 > \varepsilon_1$, equal otherwise), implying that the beating amplitude of the instantaneous displacement field intensity $|D_{tot}|^2$ oscillates between the intensity of the original incoming wave and the total sum of the forward and backward waves. As the rise time increases, the phase difference is smaller, and the standing wave ratio is correspondingly reduced.

Phenomenologically, the coupling to negative frequencies is controlled by the Fourier spectrum of the permittivity variation profile, whose bandwidth needs to be large enough to bridge the frequency gap $\Delta\omega$ between the input frequency [Fig. 2(a), blue solid] and the time-reversed waves in the final medium [Fig. 2(a), red dashed]. We illustrate this in Fig. 2(b), in which we plot the analytic Fourier transform $\mathcal{F}[\frac{d\varepsilon}{dt}](\omega)$ for the three rise-times considered in Fig. 1 (see [30] for details). It is clear how the amplitude of this quantity at frequencies $\omega \simeq \Delta\omega$ (vertical green line) determines whether or not the time-reversed wave can be efficiently excited.

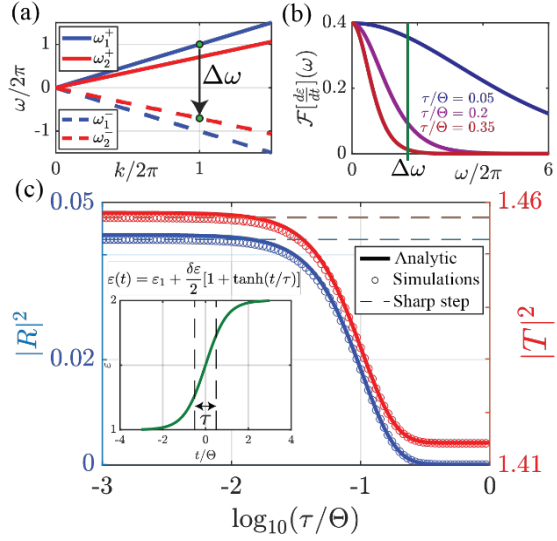

*Figure 2. (a) Dispersion relation for forward (continuous lines) and backward (dashed lines) waves before (blue) and after (red) the switching, illustrating the frequency bandwidth $\Delta\omega = \omega_1^+ - \omega_2^-$ needed for time-reversal to occur. (b) Fourier transform (for $\omega > 0$) of the derivative of the permittivity $\varepsilon(t)$ for the three switching rise-times considered in Fig. 1: time-reversal vanishes once the frequency content of the switching profile cannot bridge the frequency difference between the forward wave before the switching, and the backward wave after. (c) Scattering intensities $|R|^2$ and $|T|^2$ of the time-reversed (left axis) and transmitted (right axis) wave as a function of the switching rise time (in log-scale). Note how the difference $|T|^2 - |R|^2 = Z_1/Z_2$ is constant for any rise-time. The inset shows the sigmoidal step profile used.*

In all scenarios, independent of the transition speed, Eq. (8) for the wave amplitudes must be satisfied. As a consequence, any generated backward wave is compensated by a corresponding change in forward-propagating momentum. This can be clearly seen in Fig. 2(c), in which we plot $|R|^2$ (blue, left axis) and $|T|^2$ (red, right axis) as a function of rise time in log-scale, calculated analytically (lines) from our theory and numerically (circles) from the exact Maxwell's Equations. Indeed, the decay in amplitude of the time-reversed wave, associated with a more adiabatic temporal transition, is accompanied by a corresponding reduction in forward amplitude. The phenomenon is the dual of an adiabatic spatial interface, as in a tapered waveguide transition that suppresses unwanted reflections. Indeed, we find that the reflection coefficient rapidly converges to zero in the range in which the transition time is comparable to the period of the input wave, and for slower temporal transitions, for which no backward wave is excited, $|T(\tau \to \infty)|^2 \to Z_1/Z_2$ following Eq. (8). Details of our efficient numerical scheme used here are provided in [30]. Minor discrepancies between analytical and numerical results $< 2\%$ can be observed for larger reflections, due to the small-reflection approximation in Eq. (6).

## TEMPORAL TAPER DESIGN

Our analytical formulation yields a particularly interesting result in the isorefractive scenario, i.e., as we vary the impedance $Z_0$ but not $\gamma$. Such a scenario may be envisioned, for instance, if we vary in time the distance between two parallel plates for transverse electromagnetic wave propagation. In this case, the frequency of the wave remains constant through the temporal transition, and $\Phi(t) = 2\omega t$ in Eq. (7). Thus, assuming an arbitrary switching profile occurring from time $t = 0$ to $t = T$, we can write

$$\rho = \int_0^T F(t') e^{2i\omega t'} dt', \qquad (11)$$

which can be inverted to yield

$$F(t) = \frac{1}{\pi} \int_{-\infty}^{\infty} \rho(\omega) e^{2i\omega t} d\omega. \qquad (12)$$

Eq. (12) explicitly returns the temporal impedance profile required to synthesize a desired frequency dependence of $\rho(\omega)$ [28],[31], ideally suited, for instance, to tailor the bandwidth over which the time-reversed wave is minimized (or maximized) at will.

We can use this result to explore the optimal temporal profile that maximizes the bandwidth over which temporal reflections are suppressed for a given duration of the switching profile. The spatial analogue of this problem is known as the Klopfenstein taper [28], which describes the optimal spatial profile that maximizes the bandwidth over which reflections stay below a desired minimum value for a given taper length. In our temporal scenario, we rigorously solve this problem in [30], deriving the optimal temporal profile

$$\ln Z_0 = \frac{1}{2} \ln(Z_1 Z_2) + \frac{\rho_0}{\cosh(A)} A^2 \phi(2t/T - 1, A), \qquad (13)$$

where $A$ quantifies the bandwidth $\omega T \gtrsim A$ over which the reflection coefficient is below $\rho_{max} = \rho_0 / \cosh(A)$, and $\rho_0 = \frac{1}{2} \ln(Z_2 / Z_1)$, and the special function $\phi(x, y)$ is defined in [30]. More details on the derivation and implication of this explicit formula for the optimal switching profile can also be found in [30]. In the more general non-isorefractive scenario, we can use our general formulation to derive numerically the optimal switching profile, yielding a similar result to (13) when the index contrast is small.

Generally, our formulation enables the design of ultrafast switching profiles for optimal broadband frequency conversion with minimal back-reflections. In Fig. 3 we compare the response of a non-isorefractive temporal Klopfenstein taper (blue) with $A/2\pi T_0 \approx 3$, where $T_0 = (\varepsilon_1 \varepsilon_3)^{1/4} / 4$ is the duration of the switching profile, and maximum ripples $R_{max} \approx 0.0014$, to the one of a quarter-wave (QW) anti-reflection temporal coating (red) [26] – the temporal analogue of a conventional anti-reflection coating, consisting of two abrupt temporal interfaces delayed by a quarter period (within the middle-layer) of the lowest possible target frequency (see [30] for details). As seen in the figure, both temporal profiles feature the same initial and final permittivity $\varepsilon_1 = 1$ and $\varepsilon_2 = 2$, and same total duration, and they are both aimed at suppressing the time-reversed signals within the same frequency range. Panel (a) shows the two temporal profiles, while (b) shows the calculated $|R|$ for the QW (red) and Klopfenstein (blue) cases as a function of input frequency. The two filters are designed to work for an incoming frequency $\omega / 2\pi = 5$. In [30] we further investigate the optimal trade-off between bandwidth and reflection suppression with equal reflection peaks in the pass-band enabled by the Klopfenstein taper.

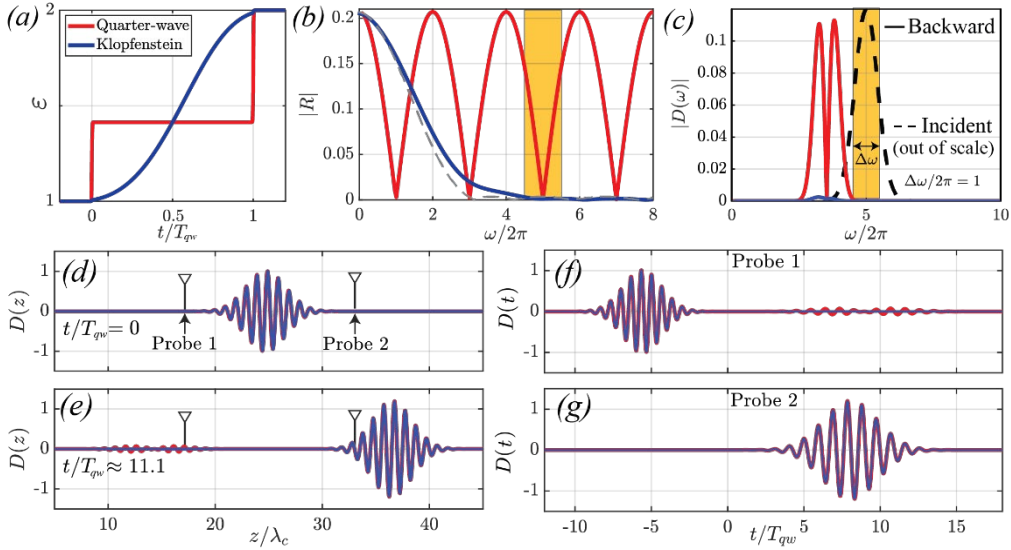

*Figure 3.* (a) Temporal permittivity profiles for a temporal quarter-wave filter (red) and a Klopfenstein taper with cutoff $A/2\pi T_{qw} = 3$, resulting in maximum ripples of amplitude $R_{max} \approx 0.0014$ (blue) and (b) their respective frequency response $|R(\omega)|$, showing the broadband response of the Klopfenstein taper, whose pass-band $\omega/2\pi \gtrsim A/2\pi T_{qw} = 3$ is characterized by ripples of constant amplitude. The dashed grey line indicates the isorefractive result corresponding to the same impedance mismatch, for which case the response has an exact zero at the pass-band edge. (c) Normalized Fourier transform of the incident (dashed) and backward-scattered (continuous) parts of a broadband pulse with carrier frequency $\omega_c/2\pi = 5$ and full width at half-maximum $\Delta\omega/2\pi = 1$ (shaded yellow region) for the switching cases: note how the Klopfenstein taper (blue) generates virtually no backward wave, whereas the quarter-wave layer cannot prevent the sidebands of the pulse from being time-reversed. (d-e) Spatial distribution of the displacement field of the pulse at times (d) $t/T_{qw} = 0$ and (e) $t/T_{qw} \approx 11.1$. (f-e) Temporal pulse profile at the locations of (f) Probe 1 and (g) Probe 2 shown in panel (d).

The difference between the QW temporal slab and the Klopfenstein taper can be appreciated when the input wave is a broadband pulse. In Fig. 3c we show the normalized spectral distribution of the incoming (dashed black line, plotted for reference out of scale) and time-reversed (continuous lines) waves computed via FDTD for a relatively broadband pulse with carrier frequency $\omega_c/2\pi = 5$ and full-width-at-half-maximum (FWHM) $\Delta\omega/2\pi = 1$ (shaded in yellow), for the two scenarios. The Klopfenstein taper produces hardly any backward wave for this pulse, leading to a pure frequency translation of $\approx 30\%$, while the QW temporal slab can only suppress time-reversal over a much narrower bandwidth around the target frequency, clearly yielding a lower efficiency. Panels (d-e) show the initial (d) and final (e) spatial field distributions for this pulse excitation, while panels (f-g) show the temporal signal at the two probes shown in panel (d), demonstrating the superior performance of the Klopfenstein taper. The employed profile is indeed optimal to maximize the bandwidth over which reflection is minimal for this time interface and its duration, as detailed in [30], where we study other taper profiles and compare their performance.

In this work, we introduced a rigorous and general analytic formulation to model temporal scattering for arbitrary switching profiles of homogeneous media, and deployed it to investigate the interplay between the finite timescale of a continuous time-switching process and the temporal variations of the impinging wave. As an application, we demonstrated how the profile of temporal switching can be tailored to control the temporal reflection in order to maximize the efficiency of time-reversal processes in ultrafast modulation setups at any frequency. Amidst the current multidisciplinary interest in exploiting mixing processes in time varying media, our findings outline the importance of considering, and possibly controlling, the switching speed and its temporal profile: the consecutive accumulation of reflection amplitude and phase throughout a continuous switching process can dictate whether time-reversal will be maximized or suppressed. Our findings set the stage for future investigations of time-reversal, frequency conversion and mixing in photonics, electromagnetics, acoustics and other wave systems undergoing temporal switching of arbitrary form, with relevant implications also in the growing area of Floquet condensed matter.

**Funding.** Simons Foundation; Air Force Office of Scientific Research; Department of Defense.

**Acknowledgments.** E.G. acknowledges the Simons Society of Fellows (855344,EG).

**Disclosures.** No disclosures needed.

**Data availability.** Data are available upon reasonable request to the authors.

**Supplemental document.** See [30] for supporting content on fast numerical solutions, Klopfenstein taper design and the associated trade-offs between bandwidth and reflection ripples, analytical details on our derivation, and comparisons between different tapering strategies.

# Supplemental document: Tapered Temporal Interfaces

## 1. Fast Numerical Solver for Continuous Time-Modulation

From Maxwell's Equations we can easily derive

$$\nabla^2 \mathbf{D} = \varepsilon(t)\mu(t)\frac{\partial^2 \mathbf{D}}{\partial t^2} + \varepsilon(t)\frac{\partial \mu}{\partial t}\frac{\partial \mathbf{D}}{\partial t}. \tag{0.1}$$

Assuming spatially harmonic solutions, we have $\nabla^2 \to -k^2$. Moreover, assuming that the modulation occurs over a finite time-interval, we have $\frac{\partial \varepsilon}{\partial t}\big|_{t \to \pm\infty} = 0$. Therefore, the solutions before and after the switching has occurred must be of the form

$$\begin{aligned} D(t_a) &= D_a^+ e^{i\omega_a t_a} + D_a^- e^{-i\omega_a t_a} \\ D(t_b) &= D_b^+ e^{i\omega_b t_b} + D_b^- e^{-i\omega_b t_b} \end{aligned}, \tag{0.2}$$

where $\omega_a$ is the initial frequency of the wave at $t = t_a$ and $\omega_b = \frac{n_a}{n_b}\omega_a$, $n_a = \sqrt{\varepsilon_a \mu_a}$, and we consider a 1D case without loss of generality. We can therefore solve Eq. (0.1) efficiently as a pair of coupled ODEs for the displacement field $D^{(0)}$ and its time-derivative $D^{(1)}$

$$\frac{\partial}{\partial t}\begin{pmatrix} D^{(0)} \\ D^{(1)} \end{pmatrix} = \begin{pmatrix} 0 & 1 \\ -\frac{k^2}{\varepsilon(t)\mu(t)} & -\varepsilon(t)\frac{\partial \mu}{\partial t} \end{pmatrix}\begin{pmatrix} D^{(0)} \\ D^{(1)} \end{pmatrix} \tag{0.3}$$

via any common ODE solver, subject to the continuity conditions for $D^{(0)}$ and $D^{(1)}$ (note that the latter condition is equivalent to the continuity of the magnetic field $B$)

$$\begin{aligned} D^{(0)}(t_j) &= e^{i\omega_j t_j} D_j^+ + e^{-i\omega_j t_j} D_j^- \\ D^{(1)}(t_j) &= i\omega_j(e^{i\omega_j t_j} D_j^+ - e^{-i\omega_j t_j} D_j^-) \end{aligned}, \tag{0.4}$$

where $j = a, b$, $D_a^+$ ($D_a^-$) is the amplitude of the forward (backward) input wave (i.e. at $t = t_a$) and $D_b^+$ ($D_b^-$) is the unknown field amplitude for the total forward (backward) wave at the end of the scattering process (i.e. at $t = t_b$).

## 2. Fourier transform of the derivative of the sigmoidal permittivity profile

The derivative of the sigmoidal permittivity profile

$$\varepsilon_s = (\varepsilon_1 + \frac{\delta\varepsilon}{2}) + \frac{\delta\varepsilon}{2}\tanh(t/\tau) \tag{0.5}$$

can be readily calculated as

$$\frac{d\varepsilon}{dt} = \frac{\delta\varepsilon}{2\tau}\text{sech}^2\{t/\tau\}, \tag{0.6}$$

and its Fourier transform reads

$$\mathcal{F}[\frac{d\varepsilon}{dt}](\omega) = \sqrt{\frac{\pi}{2}}\frac{\delta\varepsilon}{2}\text{csch}(\frac{\pi\tau\omega}{2})\tau\omega. \tag{0.7}$$

## 3. Antireflection temporal coating (temporal quarter-wave layer)

A perfect antireflection temporal coating is realized, following [1], by introducing a middle-layer with impedance $Z_{qw} = \sqrt{Z_1 Z_2}$ and duration $T_{qw} = (n_{qw}/n_1)/4$ (where $n_{qw} = \sqrt{\varepsilon_{qw}\mu_{qw}}$ is the refractive index in the middle-layer) between the two regions of time, such that the time-reversals caused by the two switching events have equal amplitude and opposite phase.

## 4. Klopfenstein taper design

Quoting Eq. (11) in the main text

$$F(t) = \frac{1}{2}\frac{\partial \ln Z_0}{\partial t} = \frac{1}{\pi}\int_{-\infty}^{\infty}\rho(\omega)e^{2i\omega t}d\omega, \tag{0.8}$$

we seek a reflection profile which enables equal-ripple amplitude and tunable bandwidth. This is known from the spatial domain to be the continuous limit of the multi-layer Chebyshev filter [1,2]:

$$\rho(\omega) = \rho_0 \frac{\cos[\sqrt{(\omega T)^2 - A^2}]}{\cosh A}e^{i\omega T}, \tag{0.9}$$

where

$$\rho_0 = \frac{Z_2 - Z_1}{Z_2 + Z_1} \simeq \frac{1}{2}\ln(\frac{Z_2}{Z_1}) \tag{0.10}$$

and the parameter $A$ determines the trade-off between the bandwidth $\omega T > A$ and the maximum ripple size through the relation

$$\rho_{max} = \frac{\rho_0}{\cosh A}. \tag{0.11}$$

Evaluating the Fourier transform of Eq. (0.9) yields the impedance profile [3]

$$\ln Z_0 = \frac{1}{2}\ln(Z_1 Z_2) + \frac{\rho}{\cosh(A)} A^2 \phi(2t/T - 1, A),  \quad (0.12)$$

where

$$\phi(t, A) = \int_0^t \frac{I_1(A\sqrt{1-y^2})}{A\sqrt{1-y^2}} dy, \quad (0.13)$$

$I_1(x)$ being the modified Bessel function. Eq. (0.13) can be evaluated efficiently by expanding $I_1(x)$ into a series and integrating term by term [4], which yields

$$\phi(t, A) = \sum_{n=0}^{\infty} a_n b_n \quad (0.14)$$

with the recursively evaluated coefficients

$$a_0 = 1;\ a_n = \frac{A^2}{4n(n+1)} a_{n-1};$$

$$b_0 = \frac{t}{2};\ b_n = \frac{\frac{t}{2}(1-t^2)^n + 2n b_{n-1}}{2n+1}. \quad (0.15)$$

## 5. Trade-off between bandwidth and reflection ripple amplitude in Klopfenstein taper

The bandwidth and the constant amplitude of the ripples in the pass-band can be traded off for each other via the parameter $A$, which also determines the maximum ripple amplitude $R_{max}$ through the relation (0.11). Note that the small-reflection assumption results in a discontinuity of the Klopfenstein impedance profile near the edges of the taper which is negligible for smaller bandwidths, but increases as $A$ is reduced, becoming dominant once the bandwidth of the taper is pushed towards lower frequencies. The taper gradually becomes equivalent to the quarter-wave layer once its bandwidth is pushed towards the first minimum of the latter, as shown in Figure S1 for four values of $A$.

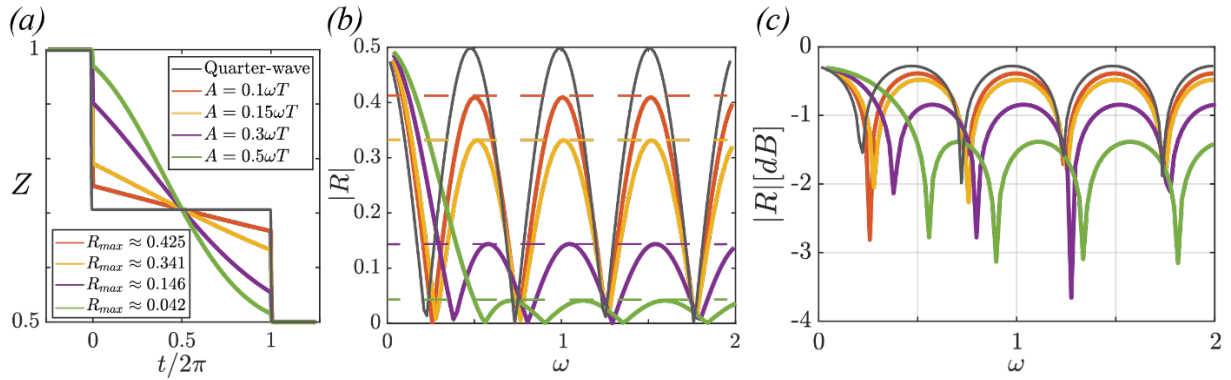

Figure S1: Dependence of the (left) Klopfenstein taper profile and (center, right) response on the bandwidth (ripple amplitude) parameter A ($R_{max}$) in linear (center panel) and logarithmic (right panel) scale.

## 6. Comparison between common tapering profiles

Common impedance tapers include the exponential and the triangular filters, as well as the optimal Klopfenstein one [1]. Figure S1 compares the response of the different filters, where the Klopfenstein filter was generated using $A/\omega_1 T = 0.75$. Note how the Klopfenstein taper provides a trade-off between bandwidth (which can be tuned via the parameter $A$) and maximum ripple amplitude in the pass band (which is $\approx 0.0062$ in this case).

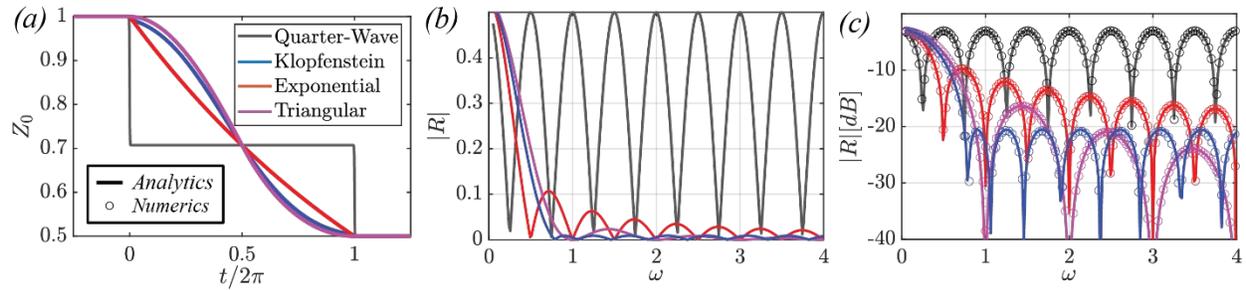

Figure S2: Comparison of temporal quarter-wave layer (grey), and Klopfenstein (blue), exponential (red) and triangular (magenta) tapers.

## 7. Non-isorefractive tapers

Although the analytical formalism for the design of temporal tapers is quantitatively more accurate in the isorefractive case, the key phenomenology presented is not exclusive to this scenario. Fig. S3 shows the performance of the different tapers for the case where the change in impedance is caused by a change in permittivity alone $\varepsilon : 1 \rightarrow 4$, such that the taper is no longer isorefractive. As for the previous scenario, in fact, the exponential taper offers a broader bandwidth, the triangular one a lower ripple amplitude (although in this case its performance is worse than the Klopfenstein filter over a broader frequency band and improves at higher frequencies) while the Klopfenstein offers an optimal tradeoff. Furthermore, note how the constant-ripple feature of the Klopfenstein taper is unaffected in the non-isorefractive case.

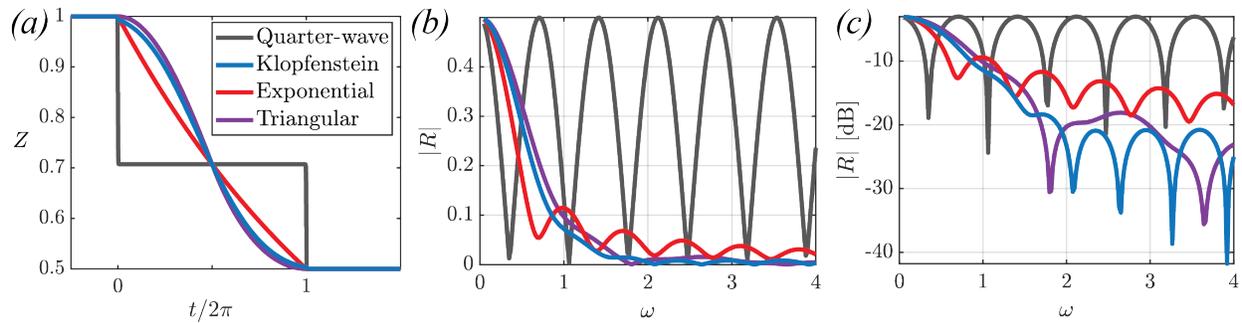

Figure S3: Comparison of different non-isorefractive temporal tapers.